\documentclass{PoS} \usepackage{wrapfig} \usepackage{amsmath}

\title{{\small{DESY 19-178, DO-TH 19/20}}\\
Resummation of large logarithms in the VFN scheme for DIS heavy-quark production
}

\ShortTitle{Large-logarithm resummation in the VFN scheme}

\author{\speaker{Sergey Alekhin}%
         \thanks{This work was supported in part by Bundesministerium f\"ur Bildung und 
                 Forschung (contract 05H18GUCC1), by EU TMR network SAGEX
                 agreement No. 764850 (Marie Sk\l{}odowska-Curie) and COST action CA16201: Unraveling new
physics at the LHC through the precision frontier.
The authors are also grateful to the Mainz Institute for Theoretical Physics 
(MITP) of the DFG Cluster of Excellence PRISMA* (Project ID 39083149) for 
its hospitality and partial support during the completion of this work.
}\\ 
II. Institut f\"ur Theoretische Physik, Universit\"at Hamburg,
    Luruper Chaussee 149, D-22761 Hamburg, Germany;\\
        Institute for High Energy Physics,142281 Protvino, Russia\\
        E-mail: \email{sergey.alekhin@desy.de}}

\author{Johannes Bl\"umlein\\ 
        Deutsches Elektronensynchrotron DESY, Platanenallee 6, D--15738 Zeuthen, Germany\\
        E-mail: \email{Johannes.Bluemlein@desy.de}}

\author{Sven-Olaf Moch\\
II. Institut f\"ur Theoretische Physik, Universit\"at Hamburg,
    Luruper Chaussee 149, D-22761 Hamburg, Germany \\      
 E-mail: \email{sven-olaf.moch@desy.de}}

\abstract {We consider the impact of the resummation of large logarithms, which appear in the QCD evolution of the heavy-quark 
distributions, on the phenomenology of deep-inelastic heavy-quark production. The heavy-quark PDFs are derived using the 
fixed-order matching conditions as a boundary for the QCD evolution and the result obtained is compared to the distributions 
defined by the matching conditions at all scales. With such an approach, the effect of heavy-quark PDF evolution is found to be 
sizable at LO and dramatically reduces at NLO. The NNLO evolved distributions are not very different from the NLO ones at large 
scales, however, show substantial differences at low virtualities, i.e. where the additional large logarithms are numerically not important, 
while a mismatch between the NLO accuracy of the matching conditions and the NNLO accuracy in the evolution kernels causes a substantial 
excess in the heavy-quark distributions. 
This excess propagates into the variable flavor number (VFN) scheme predictions for the 
deep-inelastic structure functions and has to be compensated by a decrease in the small-$x$ gluon distribution determined
from PDF fits based on the VFN scheme, which should be considered as a theoretical uncertainty in VFN PDF fits and 
reaches $\sim 30\%$ for the small-$x$ gluon distribution extracted from the data on deep-inelastic charm-quark production.  
}

\FullConference{XXVII International Workshop on Deep-Inelastic Scattering and Related Subjects (DIS2018)\\
		8-12 April 2019\\
		Turin, Italy}

\begin{document}
The heavy-quark production in deep-inelastic-scattering (DIS) of 
leptons off nucleons provides a valuable tool for the study of the 
QCD dynamics. This process can be described using the QCD-improved parton 
model with the higher-order perturbative corrections taken into account. 
In particular, a good description of the available high-precision
data on  heavy-flavor DIS production, collected in a wide range of momentum transfer $Q^2$ and 
Bjorken $x$, can be achieved by employing a factorization scheme with three light
flavors in the initial state in combination with the QCD perturbative correction up 
to the next-to-next-to-leading-order (NNLO)~\cite{Alekhin:2017kpj,Alekhin:2019ntu}. The fixed-order 
calculations still might be insufficient at extremely large $Q^2$, when the terms 
$\sim \ln (Q^2/m_h)$, where $m_h$ is the heavy-quark mass, give 
important contributions~\cite{Shifman:1977yb}. To circumvent this problem, a  
scheme with massless heavy-flavor distributions was 
suggested~\cite{Collins:1986mp}. 
These distributions are evolved similarly to the massless ones and 
in this way automatically incorporate the large logarithmic contributions through resummation
provided by the evolution equations. In the present
 study we check the impact of 
such a resummation on the QCD analysis of the available DIS data. The 
analysis framework is provided by the ABM
PDF fit, which now includes the most 
recent HERA data on the DIS $c$- and $b$-quark production~\cite{H1:2018flt}.
\begin{figure}
  \centering
  \includegraphics[width=0.7\textwidth]{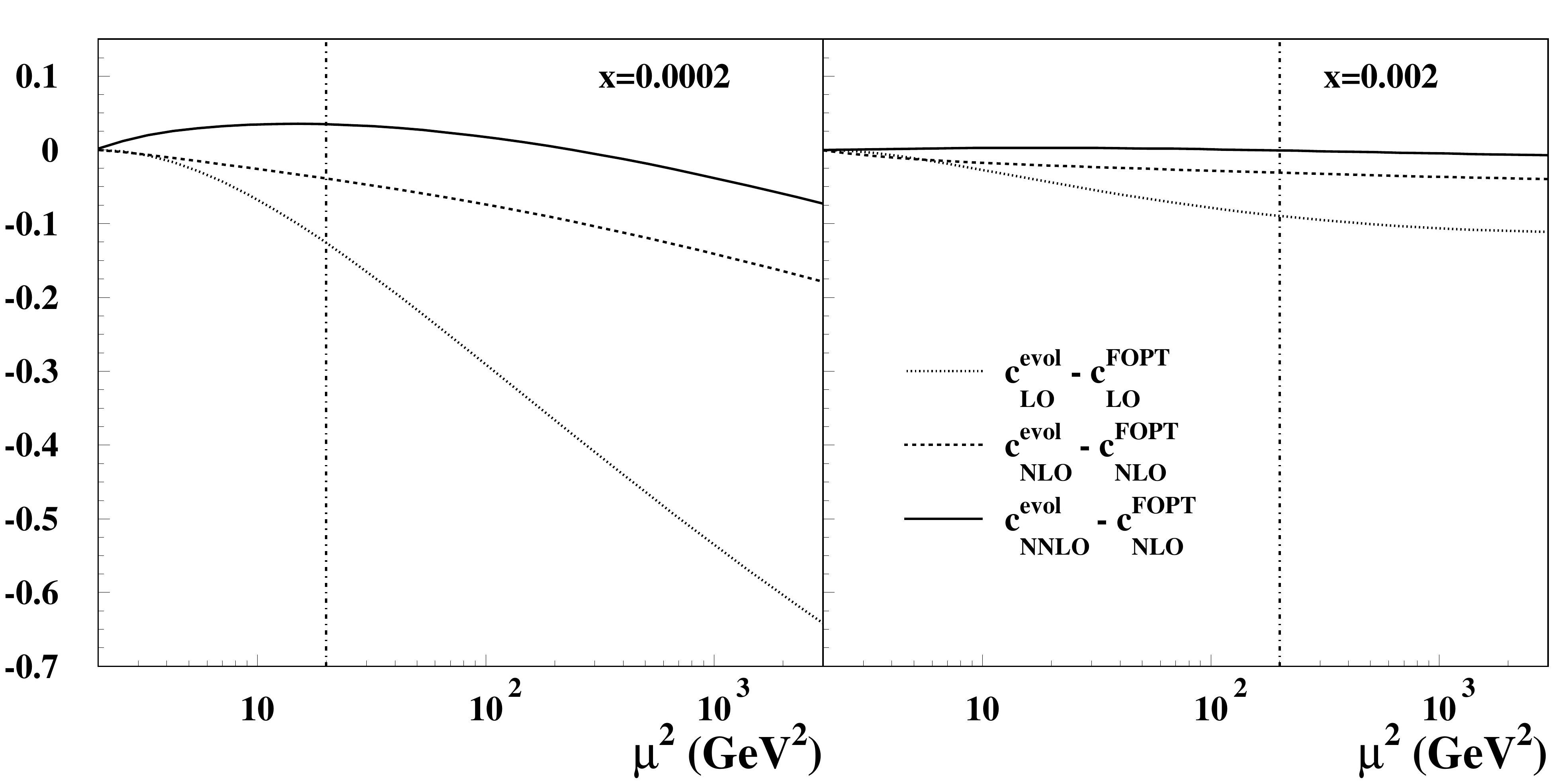}
  \caption{\small Difference between evolved $c$-quark distribution and the ones
obtained within fixed-order perturbative theory (FOPT) conditions in 
various orders of QCD (LO, NLO and NNLO)  
versus the factorization scale $\mu$ and at representative
values of the parton momentum fraction $x$ (left: $x=0.0002$, right: $x=0.002$)
taking the pole mass of $c$-quark, $m_c=1.4~{\rm GeV}$.
The vertical dashed-dotted line displays upper margin for the HERA 
collider kinematics.  
}
    \label{fig:pdfevol}
\end{figure}
These data can be well described within the 3-flavor scheme. 
However, for the purposes of the present analysis we also 
consider a variant of the ABM fit based on the massless treatment 
of the heavy-quark DIS production. There are many variants 
of the massless heavy-quark schemes available in the literature: 
ACOT, FONLL, RT and their 
numerous modifications used in various PDF fits, cf. e.g.~\cite{Bertone:2018ids}.
However, they all employ a 
common conceptual framework: The heavy-quark production 
cross sections are constructed as a combination of the terms 
corresponding to the 3-flavor and massless schemes in order to provide 
a smooth transition from the former to the latter as $Q^2$ rises. 
Such a transition is also ensured by the 
Buza-Matiounine-Smith-van Neerven (BMSN) prescription of the 
VFN scheme~\cite{Buza:1996wv,Alekhin:2009ni}.
However, the original BMSN approach is based on the 
heavy-quark PDFs derived using the fixed-order matching conditions.
Therefore resummation effects are missing in this case. 
In the present study we consider the standard BMSN prescription and its
variant with the heavy-quark PDF evolution taken into account. 
First, we compare the PDF shapes obtained within these two approaches and 
then check the impact of such a variation on the results of the global PDF
fit including the most recent HERA data on semi-inclusive $c$-quark 
DIS production~\cite{H1:2018flt}. This allows us to get a deeper insight into 
the crucial ingredients of the VFN approach and to better quantify the theoretical 
uncertainties of this formalism. 
\begin{figure}
  \centering
  \includegraphics[width=\textwidth,height=0.8\textwidth]{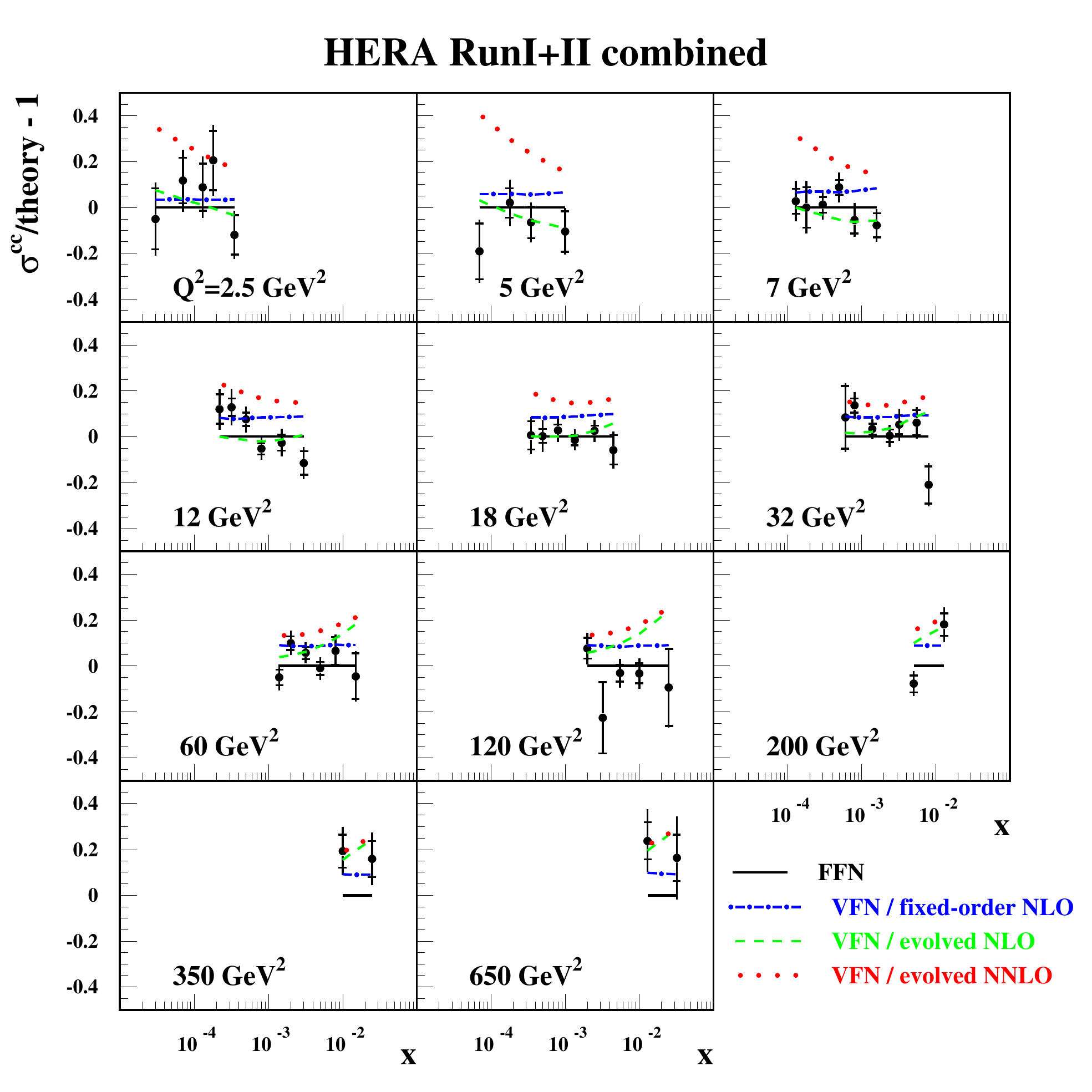}
  \caption{\small The pulls obtained for the combined HERA  data 
on DIS $c$-quark production~\cite{H1:2018flt}
in the FFN version of the present analysis (solid lines) versus $x$
in bins on $Q^2$.
The predictions obtained using the BMSN version 
of the VFN scheme with various versions of the heavy-quark PDFs with respect to the FFN 
fit are displayed for comparison (dotted-dashes: fixed order NLO, 
dashes: evolving from the 
NLO matching conditions with the NLO splitting functions, dots: the same
for the NLO matching conditions combined 
with the NNLO splitting functions).  
The 3-flavor PDFs obtained in the FFN fit
are used throughout. 
}
    \label{fig:data}
\end{figure}

The comparisons are based on the ABMP16 PDF 
fit~\cite{Alekhin:2017kpj}, which employs the fixed-flavor-number (FFN) 
scheme with three light quarks in the initial state for the description of heavy-flavor 
DIS production. An optional VFN framework is provided by
the $c$- and $b$-quark distributions, which are generated  
from the 3-flavor PDFs through  
matching conditions of Ref.~\cite{Buza:1996wv} presently 
up to the next-to-leading-order
(NLO) of perturbative QCD.\footnote{For the two-mass effects a generalized
  VFN scheme at NLO has been given in \cite{Blumlein:2018jfm}.}

In the original BMSN approach these heavy-flavor
PDFs are used for all factorization scales. In the modified BMSN scheme 
we produce the QCD-evolved PDFs using the fixed-order (FO) heavy-flavor PDFs 
as boundary condition at the initial scale, which is conventionally selected 
at the value of heavy-quark mass $m_h$, where $h=c,b$. Since the 
NLO matching is used for the boundary conditions, a consistent evolution 
implies using the NLO splitting functions. However, in order to cope with
the recent precision of DIS data the NNLO PDFs are needed. Therefore, commonly the 
NNLO evolution is used despite of the NLO accuracy of the boundary conditions. 
The evolved leading-order (LO) heavy-quark PDFs are substantially smaller than 
the fixed-order ones, cf. Fig.~\ref{fig:pdfevol}. 
For the NLO case this difference is dramatically reduced due to 
the logarithmic terms emerging in the evolution. They are partially taken into account in the NLO
terms of the FO matching conditions, although the evolution still pushes the PDFs 
to smaller values. 
For the combination of the NNLO evolution with the NLO boundary conditions
the trend changes and the evolved PDFs become larger than the FO ones.
The numerical impact of the resummation is still limited 
since even at very large scales the NNLO PDFs are not very
different from the FO and evolved ones obtained at  NLO. Instead, 
the effect manifests itself more significantly at small scales. Therefore it 
is related rather to the mismatch between the theoretical accuracy 
in the boundary conditions and the evolution kernel.
This discrepancy should disappear with
account of the upcoming NNLO corrections to OMEs~\cite{HQ}. 
Meanwhile, the difference between the FO NLO and evolved NNLO PDFs 
provides an estimate of the theoretical uncertainty
in the NNLO implementation of VFN schemes.

To check the impact of this uncertainty on the PDFs obtained from 
analysis of the DIS data we perform 
several variants of the PDF fit with different treatments of 
the heavy-quark contributions. 
For a better discrimination of the schemes we employ in the fit a recent
combination of the H1 and ZEUS data on the DIS $c$- and $b$-quark  
production with reduced uncertainties~\cite{H1:2018flt}.
Furthermore, we drop the inclusive HERA data in order to show 
the sensitivity of the PDFs to the scheme choice in greater detail. 
For the same reason, the collider data on $W$- and $Z$-boson production, 
which provide some constraint on the small-$x$ gluon distribution, are also 
dropped and the data on DIS off deutron targets are added instead in 
order to keep disentangling of the $u$- and $d$-quark distributions.
The FFN fit is performed taking the NLO massive Wilson coefficients with
the pole-mass definition~\footnote{The transition
to the heavy-quark mass definition in the $\overline{\rm MS}$ scheme for the 
corresponding heavy-flavor Wilson coefficients is known, cf.~\cite{HQ}.}
 and the value of $m_c^{pole}=1.4~{\rm~GeV}$,
which ensures a good description of the HERA data, cf. Fig.~\ref{fig:data}.
This choice is also needed for a consistent comparison with the version of 
the VFN scheme based on the NLO heavy-quark evolution, given 
available theoretical accuracy of the massive OMEs. 

\begin{figure}
  \centering
  \includegraphics[width=0.455\textwidth,height=0.42\textwidth]{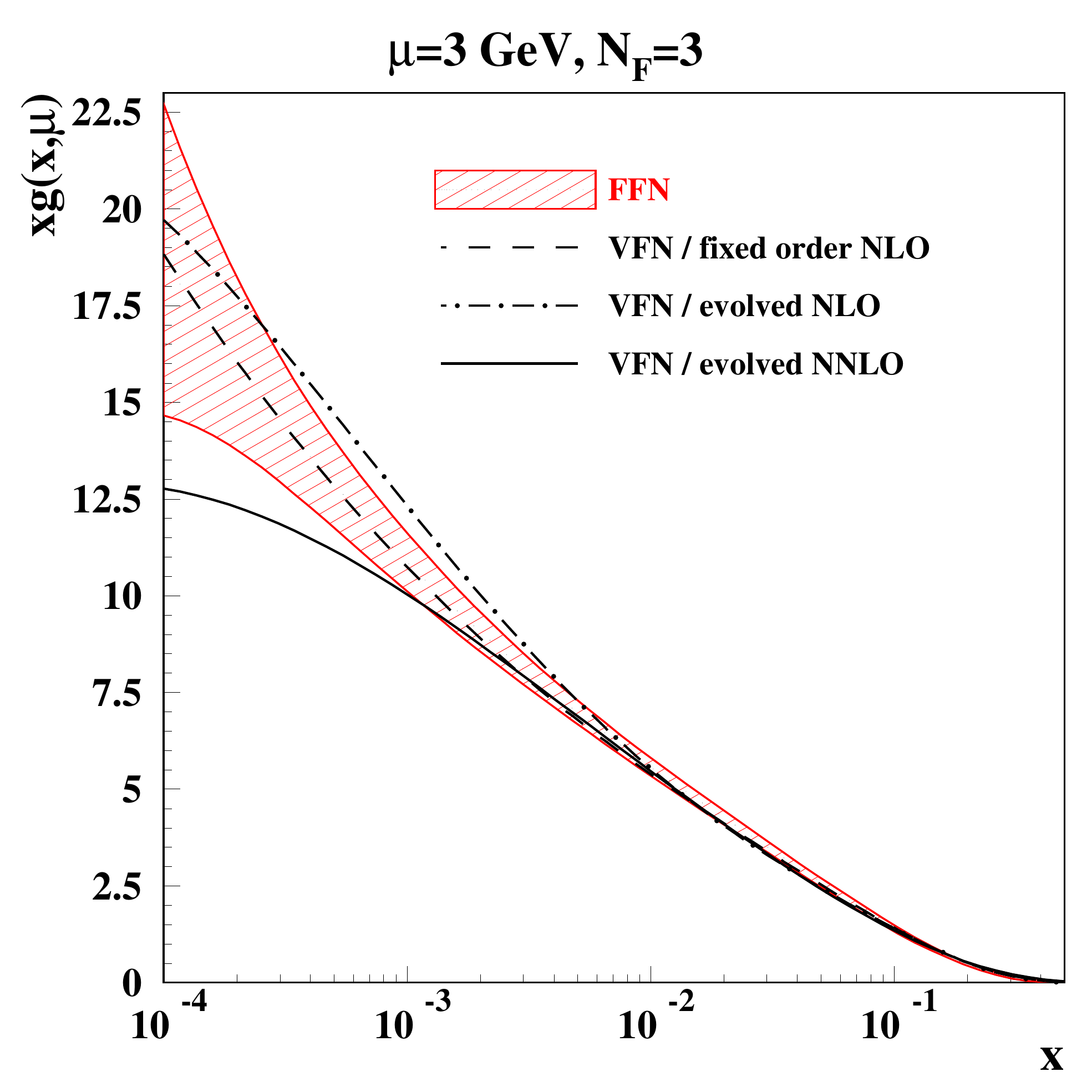}
  \includegraphics[width=0.455\textwidth,height=0.42\textwidth]{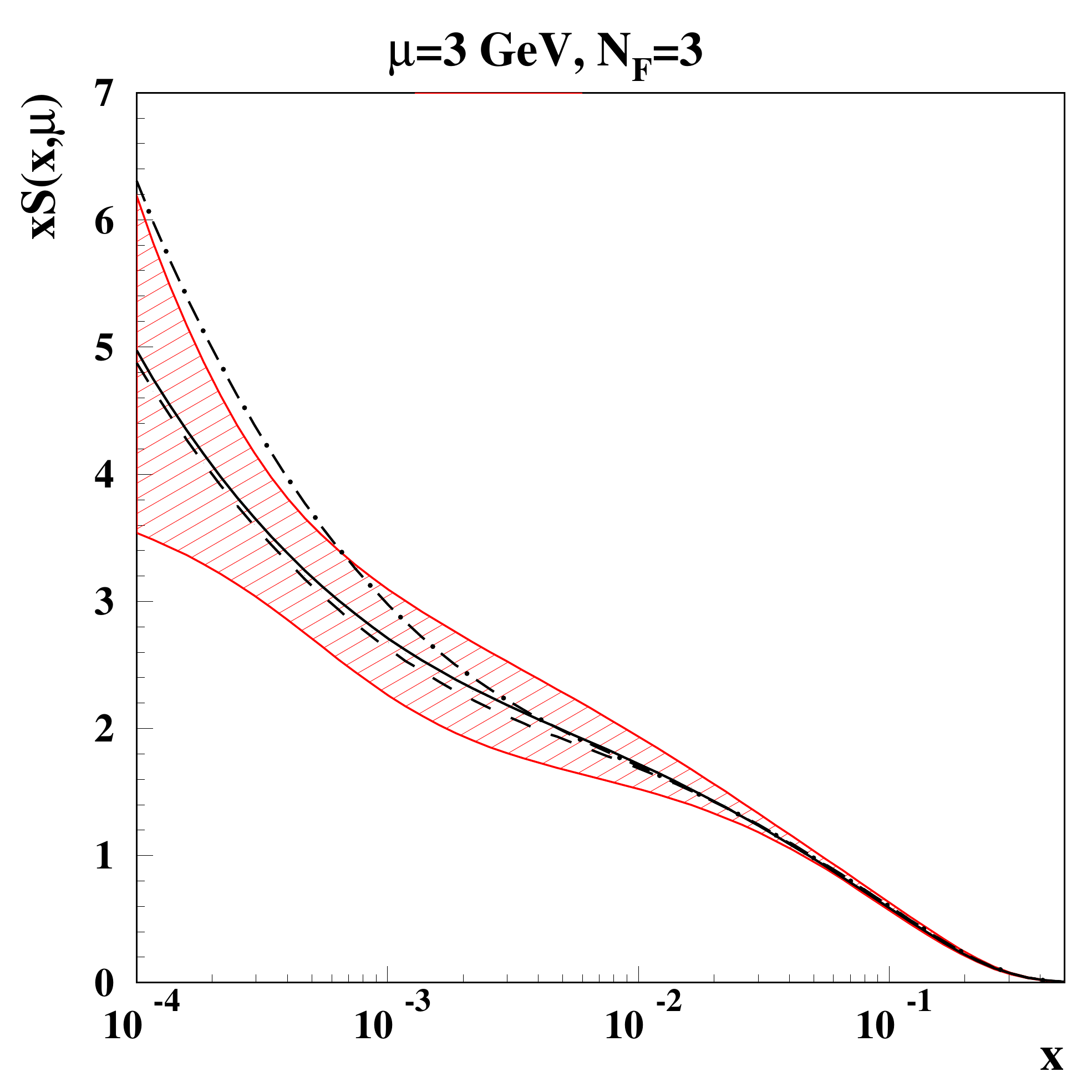}
  \caption{\small The central values of 3-flavor gluon $xg(x,\mu)$ 
(left) and the total 
light-flavor sea $xS(x,\mu)$ (right) distributions 
at the factorization scale $\mu=3~{\rm GeV}$ versus $x$ obtained 
in the various versions of the VFN scheme (solid: evolved NNLO, 
dashes: fixed-order NLO, dashed-dots: evolved NLO). The $1\sigma$
error band for the FFN results is given for comparison (hatched area). 
}
    \label{fig:pdfs}
\end{figure}

The present analysis employs a BMSN prescription of the VFN scheme, which 
reads for the $c$-quark production structure function $F_{2,c}$ as follows:
\begin{equation}
F_{2,c}^{\rm BMSN}(x,Q^2)=F_{2,c}^{\rm FFN}(x,Q^2)+F_{2,c}^{\rm ZMVFN}(x,Q^2)-F_{2,c}^{\rm ASYM}(x,Q^2),
\label{eq:bmsn}
\end{equation}
where $F_{2,c}^{\rm FFN}$ is the expression for the 3-flavor scheme with massive 
Wilson coefficients,  $F_{2,c}^{\rm ASYM}$ describes the limit of $F_{2,c}^{\rm FFN}$ for 
asymptotic values of $Q^2 \gg m_c^2$ and
$F_{2,c}^{\rm ZMVFN}$ is computed using the 4-flavor PDFs 
in combination with the massless Wilson coefficients. 
At large values of $Q^2$  $F_{2,c}^{\rm FFN} \approx F_{2,c}^{\rm ASYM}$. 
Therefore $F_{2,c}^{\rm BMSN}(x,Q^2)$ reproduces 4-flavor VFN scheme expression.
At low values of $Q^2$ $F_{2,c}^{\rm ZMVFN}\approx F_{2,c}^{\rm ASYM}$ and 
$F_{2,c}^{\rm BMSN}\approx F_{2,c}^{\rm FFN}$.
Furthermore, a smooth transition between FFN and VFN schemes is provided 
when the FO $c$-quark distributions are employed in 
$F_{2,c}^{\rm ZMVFN}$~\cite{Alekhin:2009ni}. For this reason the BMSN predictions
obtained with the FO heavy-flavor PDFs are similar to the FFN ones for the kinematics
of existing data, cf. Fig.~\ref{fig:data}. The BMSN predictions 
made with the NLO-evolved PDFs are close to the FO results, in line with the 
comparison of Fig.~\ref{fig:pdfevol}. Meanwhile, the use of NNLO-evolved 
PDFs leads to much larger values, particularly at low $Q^2$. 
Due to this excess the results obtained with the VFN scheme based on the 
heavy-quark NNLO evolution are substantially different from the ones of 
other fit variants. The small-$x$ gluon 
distribution is most sensitive to details of the VFN scheme
settings, cf. Fig.~\ref{fig:pdfs}, and the spread observed gives an estimate of 
the theoretical scheme uncertainties due to missing higher orders of perturbative QCD.  
In particular, the difference which appears
due to switching between the NLO- and NNLO-evolution ansatz, 
related to the yet incomplete NNLO corrections in the massive OMEs, is more sizable
for the small-$x$ gluon distribution and amounts to $\sim 30\%$ at $x\sim 10^{-4}$. 

It is worth to mention that this uncertainty is relevant for all 
versions of the VFN schemes employed in PDF fits. Meanwhile, the mismatch 
between perturbative accuracy of the massive OMEs and the massless 
evolution kernels also leads to the nonphysical kink in $Q^2$-dependence of 
the DIS structure functions at $Q^2~\sim m_h^2$. 
In many VFN scheme implementations this kink is 
smoothened out by introducing various damping coefficients, 
which also formally reduce 
corresponding theoretical uncertainty. Since the shape of 
such coefficients 
is not based on solid theoretical arguments, they in turn, 
introduce additional uncertainties~\cite{Gao:2013wwa}, 
which effectively reflect the original one. 
For the FFN description of the DIS data this 
uncertainty is irrelevant, but it appears in the PDF fits including 
collider data on the production of massive final states (e.g., $W$-, $Z$-, Higgs-bosons), 
which are routinely described using the 5-flavor PDFs. Its impact 
is reduced as compared to the VFN fits due to the fact that it is localized at small factorization 
scales, cf. Fig.~\ref{fig:pdfevol}. However,  
the complete NNLO OMEs are still required 
in order to achieve ultimate PDF precision~\cite{HQ}. 

In summary, we considered the impact of the resummation of large logarithms, which 
appear in the QCD evolution of the heavy-quark distributions, on the phenomenology 
of the DIS heavy-quark production. 
The heavy-quark PDFs are derived using the fixed-order matching conditions
as a boundary for the QCD evolution and the result obtained 
has been compared to the distribution defined by the matching conditions 
at all scales. 
With such an approach the effect of heavy-quark PDF evolution 
is found to be sizable at LO and to be dramatically reduced at NLO, 
since the large logarithmic terms, which are resumed by the evolution, 
are partially taken into account 
in the NLO corrections to the fixed-order distributions. 
The NNLO evolved distributions are not very different from the NLO ones
at large scales, however, demonstrate a substantial difference at low values of $Q^2$.
This implies, that the additional large logarithms are numerically unimportant, while 
a mismatch between the NLO accuracy of the matching conditions and 
NNLO accuracy in the evolution kernels results into a substantial excess of
the heavy-quark distributions at small scales. 
This excess propagates into the variable flavor number (VFN) scheme predictions for the 
deep-inelastic structure functions and has to be compensated by a decrease in the small-$x$ gluon distribution determined
from PDF fits based on the VFN scheme, which should be considered as a theoretical uncertainty in VFN PDF fits and 
reaches $\sim 30\%$ for the small-$x$ gluon distribution extracted from the data on deep-inelastic charm-quark production.  


\end{document}